\documentclass[11pt,twoside]{article}
\usepackage{asp2014,color,natbib}
\usepackage[version=4]{mhchem}

\aspSuppressVolSlug
\resetcounters

\begin{document}

\title{Observing the Effects of Chemistry on Exoplanets and Planet Formation}
\author{Brett A. McGuire$^1$, Edwin Bergin$^2$, Geoffrey A. Blake$^{3,4}$, Andrew M. Burkhardt$^5$, L. Ilsedore Cleeves$^6$, Ryan A. Loomis$^1$, Anthony J. Remijan$^1$, Christopher N. Shingledecker$^{7,8}$, and Eric R. Willis$^9$}
\affil{$^1$National Radio Astronomy Observatory, Charlottesville, VA, 22903; \email{bmcguire@nrao.edu}}
\affil{$^2$Department of Astronomy, University of Michigan,  Ann Arbor, MI 48109}
\affil{$^3$Division of Geological and Planetary Sciences, California Institute of Technology, Pasadena, CA 91125}
\affil{$^4$Division of Chemistry and Chemical Engineering, California Institute of Technology, Pasadena, CA 91125}
\affil{$^5$Harvard-Smithsonian Center for Astrophysics, Cambridge, MA 02138}
\affil{$^6$Department of Astronomy, University of Virginia, Charlottesville, VA 22904}
\affil{$^7$Max Planck Institute for Extraterrestrial Physics, Garching, Germany}
\affil{$^8$Institute for Theoretical Chemistry, University of Stuttgart, Germany}
\affil{$^9$Department of Chemistry, University of Virginia, Charlottesville, VA 22904}

\paperauthor{Brett A. McGuire}{bmcguire@nrao.edu}{0000-0003-1254-4817}{National Radio Astronomy Observatory}{}{Charlottesville}{VA}{22903}{USA}
\paperauthor{Edwin Bergin}{}{}{University of Michigan}{}{Ann Arbor}{MI}{48109}{USA}
\paperauthor{Geoffrey A. Blake}{}{}{California Institute of Technology}{}{Pasadena}{CA}{91125}{USA}
\paperauthor{Andrew M. Burkhardt}{}{}{Harvard-Smithsonian Center for Astrophysics}{}{Cambridge}{MA}{02138}{USA}
\paperauthor{L. Ilsedore Cleeves}{}{}{University of Virginia}{}{Charlottesville}{VA}{22904}{USA}
\paperauthor{Ryan A. Loomis}{}{}{National Radio Astronomy Observatory}{}{Charlottesville}{VA}{22903}{USA}
\paperauthor{Anthony J. Remijan}{}{}{National Radio Astronomy Observatory}{}{Charlottesville}{VA}{22903}{USA}
\paperauthor{Christopher N. Shingledecker}{}{}{Max Planck Institute for Extraterrestrial Physics}{}{Garching}{}{}{Germany}
\paperauthor{Eric R. Willis}{}{}{University of Virginia}{}{Charlottesville}{VA}{22904}{USA}



\vspace{1em}

\section{Background}

The location of condensation fronts (snow lines) in protoplanetary disks (PPDs) can have substantial effects on the physical and chemical evolution of planets in the system.  Indeed, many models predict that these condensation fronts are the locations where planetesimal formation begins \citep{Ros:2013ey,Schoonenberg:2017do,Drazkowska:2017gq}.  The location of these condensation fronts is determined both by the incident radiation from the protostar, but also, in large part, by the chemical composition of the gas and ice.  The desorption temperature of the bulk ice varies significantly with the chemical composition and mixing ratio \citep{Collings:2004jp}, even of minor constituents.  This can be further complicated by a variance in the sticking efficiency of the icy grains with composition \citep{Wang:2005jm}.  It is clear that the underlying chemical composition in these ices, even at low mixing ratios, can have a substantial effect on the efficiency of planet formation.\\\vspace{-0.6em}

Further, the relative gas vs solid-phase inventories at snow lines govern the composition of the planets and their atmospheres.  These compositions (e.g., the C/O ratio in planetary atmospheres; \citealt{Oberg:2011je}) may therefore be useful probes of the history of exoplanets.  Effectively using observed exoplanet elemental ratios to test those models therefore relies heavily on our ability to accurately interpret chemical abundances measured in disks to infer the inventory of undetectable species, and how these measurements affect observed elemental ratios. \\\vspace{-0.6em}

Constraining the chemical compositions in PPDs is challenging.  The most complex gas-phase species yet detected in PPDs are methanol (\ce{CH3OH}; \citealt{Walsh:2016bq}) and methyl cyanide (\ce{CH3CN}; \citealt{Oberg:2015dh}), and more complex molecules are unlikley to be observed \citep{Walsh:2014jq}. Further, in cold or shielded regions within disks, ices may be the major repository of C and O \citep{Bergin:2016ge}. Direct observation of ices typically require space-based infrared facilities and a target source with a sufficiently bright background continuum \emph{and} favorable viewing geometry.   Even in massive young stellar objects with substantial ice columns, the list of firmly-detected ice species numbers less than a dozen \citep{Boogert:2015fx}.   Our knowledge of gas and ice compositions in disks must therefore rely on astrochemical models, using the detectable simple gas species as the primary constraint.\\\vspace{-0.6em}

These models are not yet robust, despite substantial efforts \citep{Garrod:2013id,Herbst:2017hg,Wakelam:2015dr}.  To reliably infer chemical compositions in PPDs, in which chemistry is closely coupled to physical conditions (temperature, density, UV flux, evolving dust populations, etc.) that vary with radius and scale height, necessitates that these models be well-constrained by observations in earlier stages of star formation, where more robust inventories can be observed. Here, we briefly discuss the state-of-the-art in progressively earlier stages of chemical evolution, and outline future directions and the technology and community support needed to enable them.\\\vspace{-0.6em}

\section{Chemical evolution in PPDs ...} 

A number of larger scale studies of disk physical properties and chemistry have been enabled by the sensitivity and resolution of ALMA.  Efforts to measure fundamental properties such as bulk mass \citep{Ansdell:2016gm}, and larger compositional studies examining nitrogen \citep{Guzman:2017ek} and deuterium \citep{Huang:2017dy} ratios, are now possible for larger samples sizes. With robust chemical modeling, ALMA observations of simple hydrocarbons (e.g., \ce{C2H}) can perhaps be used to infer C/O ratios indirectly \citep{Bergin:2016ge}.  As well, the JWST and, if commissioned, Origins Space Telescope, have significant potential to observe major ice components (e.g. \ce{H2O}, CO, \ce{CO2}, \ce{CH3OH}) toward a wide variety of targets.  Yet, these infrared facilities will still largely be blind to the complex chemistry occurring within the ices, and to the combined ice/gas coupling in warmer gas.  That insight is only achievable through radio observations of gas-phase species, that are sensitive to populations orders of magnitude lower in abundance, coupled with chemical modeling.\\\vspace{-0.6em}

\section{... and in complex, star-forming regions} 

Sgr B2(N), Orion-KL, and NGC 6334I are the most chemically and physically complex stages of star formation observed and modeled \citep{Garrod:2008tk,Crockett:2014er,McGuire:2017gy}.  The extreme physical environments, interplay between the gas and ice, and complex structures often result in models which can reproduce observed abundances in, at best, a qualitative fashion \citep{Laas:2011yd,Garrod:2008tk,Garrod:2013id}.  These types of physical and chemical interactions are the same that are at play in PPDs, albeit on a much larger scale.  It is therefore these environments that provide a unique test environment for the similar PDR style models applied to PPDs.  Improvements in these models are most likely to be achieved by testing them against a wide range of environments from quiescent clouds to massive star forming regions.\\\vspace{-0.6em}

Existing facilities in the (sub-)millimeter regime, particularly ALMA, are capable of reaching the line-confusion limit in broadband line surveys of complex, star forming regions in a matter of hours.  That is, the density of molecular lines in the resulting spectra is such that there is at least one spectral line every FWHM; there are no line-free channels.  When line-confusion is reached, no additional information can be gained from longer integration times at the same spatial resolution.  We are rapidly approaching the point at which deep drill observations with ALMA and the GBT will no longer produce new spectral lines.  The end result of this limit is that no further detections of new species can be made, and thus the refinement of chemical networks will be limited to those species detectable at the confusion limit.\\\vspace{-0.6em}

In the cm-wave regime, however, the most sensitive survey, the Prebiotic Interstellar Molecular Survey (PRIMOS; \citealt{Neill:2012fr}) with the most sensitive facility (the GBT) toward the most line-rich source (Sgr B2) has lines in only $\sim$10-20\% of its channels at cm-wavelengths.  Diminishing returns on sensitivity for this survey have already set in, and to increase the sensitivity substantially would require hundreds of hours of dedicated observations for every $\sim$2 GHz window.  Further, ALMA observations at higher frequencies are revealing that the region displays widely variable physical conditions within a relatively small environment, variation that is lost to single-dish GBT observations \citep{Belloche:2016fm}.  This offers a valuable opportunity to probe environment-induced chemical differentiation, but only with spatially-resolved observations which are not possible with the GBT.\\\vspace{-0.6em}

While the brightest lines in Sgr B2 are readily detectable with extant facilities such as the VLA and ATCA \citep{Corby:2015bv}, the complex species are not efficiently observable, especially at high angular resolution.  The only viable solution to this problem is the construction of a next-generation interferometer like the next-generation Very Large Array (ngVLA).  Under the current design concept \citep{Selina:2017wb}, the ngVLA is likely to not only be capable of spatially resolving chemistry in Sgr B2(N), but to reach the line-confusion limit even at cm-wavelengths, providing substantial new, complex molecular detections and their correlation with the physical conditions at work.\\\vspace{-0.6em}

\section{Applications to PPDs}  

Once refined, these models can be used to accurately interpret observations from ALMA targeting the outermost condensation fronts within disks, primarily CO snow lines \citep{Qi:2015dx}.  Application in the terrestrial planet forming region, near the \ce{H2O} snow line, will require AU-scale resolution with high sensitivity.  A high-frequency ngVLA is the only facility currently in design with the required specifications.  Further, while these models can then interpret results for C and O in disks, observations of the major N-bearing species, \ce{NH3} will still be required at similar scales.  An ngVLA operating at least through the 24~GHz \ce{NH3} transition is therefore necessary.\\\vspace{-0.6em}

An ngVLA design concept is needed that supports broadband, high surface-brightness sensitivity capabilities at the small to modest (0.1--3\,$^{\prime\prime}$) scales required to probe the effects of the widely variable physical conditions on the chemical evolution in sources like Sgr B2 and unlock chemical inventories obscured by line confusion suffered at extant facilities like ALMA.  The same instrument should be capable of observing \ce{NH3} and imaging the planet forming zone inside the water snow line in disks, requiring AU-scale (mas) resolution capabilities.\\\vspace{-0.6em}

\section{Chemical evolution in protostellar environments} 

The protostellar stage of star formation spans a wide range of sources, from `hot corinos' such as IRAS 16293 \citep{Jorgensen:2016cq} to chemically rich outflow sources like L1157 \citep{Mendoza:2014kna}. The molecular outflows from active young protostars like L1157 provide a unique environment to test the immediate impact of rapid changes in temperature and density on `pristine' molecular clouds. In these sources, distance from the protostar along the outflow corresponds directly to time, allowing modeling of temporal evolution.  Hot corinos are the next stage after L1157 and before hot cores like Sgr B2(N).\\\vspace{-0.6em}

The best method for constraining the chemistry in these sources is broadband spectral line surveys.  These sources are relatively compact, are beam-diluted in single-dish observations, and their linewidths are typically such that high-sensitivity data are often at or approaching line confusion limits, necessitating interferometric observation \citep{Jorgensen:2016cq}.  Because they are typically moderately warm (50--150~K; \citealt{Jorgensen:2016cq,McGuire:2015bp}), complex molecules display bright features in the 1--3~mm window well-covered by ALMA, NOEMA, and the SMA.\\\vspace{-0.6em}

The SOLIS project at NOEMA and IRAM, which examines a sample of solar-type protostellar sources \citep{Ceccarelli:2017ff}, and the PILS project at ALMA, which has focused particularly on an exquisitely detailed examination of the IRAS 16293  system \citep{Jorgensen:2016cq}, are the primary active dedicated investigations into this class of objects.  Extant observational facilities are likely sufficient to provide the required insight.  Ongoing and future efforts on ALMA, NOEMA, and the SMA examining a large sample of protostellar environments are critical to constraining the effects of physical environment on chemical evolution and incorporating these effects into models.\\\vspace{-0.6em}


\section{Chemical evolution in quiescent environments}  

Cold (T$_{ex}$ $\sim$10~K), dark, starless clouds are an ideal environment in which to test the time evolution of chemical reaction networks under `constant' physical conditions.  The typical sources for exploring this stage of star and planet formation are highly extended ($\theta _s$~$>$~30$^{\prime\prime}$).  At these excitation temperatures, most molecules, especially those more complex than CH$_3$OH, will display bright emission at cm wavelengths.  The best strategy for obtaining the required abundance and excitation information is therefore deep, broadband spectral line surveys at cm-wavelengths using single-dish facilities combined with chemical modeling \citep{Cordiner:2017dq,Kaifu:2004tk,McGuire:2018it}.  The best extant instrument for this work is the Green Bank Telescope (GBT).\\\vspace{-0.6em}

Even in a source as extended as TMC-1, there exists evidence for large chemical variation over small spatial scales (0.02--0.03~pc; \citealt{Peng:1998jm,Dickens:2001fo}), and observations of such would probe the effects of the very earliest stages of collapse on the chemistry. These observations will require high spatial resolution, high surface-brightness sensitivity, and broad spectral bandwidth to be efficient. Because of the cold temperatures, this further requires a cm-wavelength facility.  While the VLA satisfies this need, it lacks the combination of surface brightness sensitivity and bandwidth necessary to conduct a sensitive line survey with any efficiency. The proposed ngVLA, in its current conceptual design, would prove more than optimal for this task.\\\vspace{-0.6em}

The chemical evolution of all ranges of environments from planetary atmospheres \citep{Arsenovic:2016bc} to disk midplanes \citep{Cleeves:2013cc} to molecular clouds \citep{Hudson:2008hf} is impacted by radiation. The last few years have seen significant advances in modeling efforts to accurately include these effects \citep{Shingledecker:2018fm}, but these efforts are young, and require substantial training on `simple' environments like quiescent clouds where the radiation effects can be easily decoupled from dynamical physical events such as outflows and shocks.  In short, efforts at the GBT and other large single-dish facilities are needed to survey quiescent chemical clouds.  These efforts would be greatly bolstered by the development of broadband capabilities on the ngVLA to enable investigations of the chemical evolution upon initial gravitational collapse.\\\vspace{-1.5em}

\begin{center}
\large
    \textbf{Summary}
\end{center}

A combination of observational and theoretical efforts are needed that are focused on placing rigorous constraints on models of chemical evolution at each stage of the star and planet formation process.  The observational requirements of these investigations necessitate a commitment by the community to support extant facilities and the development of technological advancements for the same, as well as the development of a next-generation VLA with high sensitivity and exquisite spatial resolution.  The ultimate goal should be producing models that can be used to better calibrate our estimates of gas and ice compositions in PPDs, given the observations of simple gas-phase species achievable in these sources, and the construction of a radio facility capable of observationally exploiting these models \emph{in situ} during planet formation.\\

\end{document}